\documentclass[twocolumn,prb]{revtex4}
\usepackage{amsfonts, amsmath, latexsym}
\usepackage[dvips]{graphicx}
\usepackage{fancyhdr}

\begin{document}

\title{
Solvent coarse-graining and the string method\\ applied to the hydrophobic collapse of a hydrated chain
  } 
\author{Thomas F. Miller, III} \email{tfmiller@berkeley.edu}
\author{Eric Vanden-Eijnden} \email{eve2@cims.nyu.edu}
\author{David Chandler}
\email{chandler@cchem.berkeley.edu} 
 \affiliation{$^{*,\ddag}$Department of
  Chemistry, University of California, Berkeley, CA 94720}
  \affiliation{$^\dag$Courant Institute of Mathematical Sciences, New York
  University, New York, NY 10012}

\date{\today}

\begin{abstract}
Using computer simulations of over 100,000 atoms, the mechanism for the hydrophobic collapse of an idealized hydrated chain is obtained.  This is done by coarse-graining the atomistic water molecule positions over 
129,000 collective variables that represent the water density field
and then using the string method in these variables to compute the minimum free energy pathway (MFEP) for the collapsing chain.
The dynamical relevance of the MFEP (i.e. its coincidence with the mechanism of collapse) is validated \textit{a posteriori} using conventional molecular dynamics trajectories.
Analysis of the MFEP provides atomistic confirmation for the mechanism of hydrophobic collapse proposed by ten Wolde and Chandler.
In particular, it is shown that lengthscale-dependent hydrophobic dewetting is the rate-limiting  step in the hydrophobic collapse of the considered chain. 
\end{abstract}

\maketitle
 
 








\section{Introduction}
This paper applies the string method\cite{E02,E05,Mar06} to the phenomenon of hydrophobic collapse.
It is shown that the method can describe complex dynamics in large atomistic systems, ones for which other currently available rare event methods would seem intractable.
Furthermore, the string method is used to demonstrate that atomistic dynamics can be usefully projected onto that of a coarse-grained field.
The specific application of the string method considered herein finds results that are consistent with the mechanism of hydrophobic collapse put forward by ten Wolde and Chandler.\cite{ten02}

The hydrophobic effect - or the tendency of oil and water to separate
on mesoscopic lengthscales - has long been recognized as an important
driving force in molecular assembly.\cite{Kau59} It stabilizes the
formation of micelles, protein tertiary structure, multi-protein
assemblies, and cellular membranes.\cite{Saf94,Tan78} Recent
theoretical developments have helped establish a quantitative
understanding of the thermodynamics of hydrophobicity,\cite{Cha05,Hum00} but
the dynamics of hydrophobic collapse remain poorly understood because
it couples a large range of length- and timescales.
Relevant processes
include the atomic-scale motions of individual water molecules,
collective solvent density fluctuations, and the
nanometer-scale movements of the hydrophobic solutes.  Bridging these
dynamical hierarchies - and addressing the problem of complex dynamics
in large systems - is a fundamental challenge for computational
methods.

We address this challenge using the string method in collective variables.\cite{E02,E05,Mar06}
We consider the collapse of a chain composed of twelve spherical hydrophobes in an explicit solvent of approximately 34,000 water molecules. 
The system is studied by coarse-graining the water molecule positions onto a 
set of 129,000 collective variables that represent the solvent density field and then using the string method in these variables to compute the minimum free energy path (MFEP) for the hydrophobic collapse of the chain.
Conventional molecular dynamics (MD) simulations are subsequently performed to confirm that this coarse-grained description adequately describes the mechanism of hydrophobic collapse.

ten Wolde and Chandler\cite{ten02} have previously reported simulations of a non-atomistic model for the hydrated chain considered herein.
It was found that the key step in the collapse dynamics is a collective solvent density
fluctuation  that is nucleated at
the hydrophobic surface of the chain.  However, it was not
clear whether this proposed mechanism captured the essential features
of hydrophobic collapse or whether it was an artifact of the model.  Atomistic simulations were needed
to resolve the issue.

Previous efforts to characterize the mechanism of hydrophobic collapse
using atomistic computer simulations neither confirm nor disprove the
mechanism proposed by ten Wolde and Chandler.  In recent work, for
example, molecular dynamics (MD) simulations showed that dewetting
accompanies the collapse of hydrophobes in
water.\cite{Hua03,Hua04,Liu05,Ath07} However, the rate-limiting step - and
thus the mechanism for hydrophobic collapse - was not characterized.
Specifically, in Ref. \onlinecite{Liu05}, MD trajectories were
initiated at various separation distances for a pair of melettin
protein dimers.  Observation of the collapse dynamics was observed
only when the initial configuration was on the ``near'' side of the
free energy barrier, but the actual nature of that barrier - and the
dynamics of crossing it - were not studied.  In Ref. \onlinecite{Ath07}, the thermodynamics and solvation of a hydrophobic chain was studied as a function of its radius of gyration, but again, the dynamics of collapse was not characterized.

The results presented here are the first atomistic simulations to explicitly confirm the mechanism of hydrophobic collapse put forward by ten Wolde and Chandler.\cite{ten02} 
In particular, we show that the rate-limiting step in hydrophobic collapse coincides with a collective solvent motion and that performing the rate-limiting step involves performing work almost exclusively in the solvent coordinates.
We further show that the solvation free energy along the MFEP can be decomposed into small- and large-lengthscale contributions.  This analysis demonstrates that atomistic solvent energetics can be quantitatively modeled using a grid-based solvent density field, and it suggests that the rate-limiting step for hydrophobic collapse coincides with lengthscale-dependent hydrophobic dewetting.

Combined, the results presented in this study provide (1) evidence that the string method is a powerful tool for atomistically simulating complex dynamics in large systems, (2) a proof of principle that atomistic solvent dynamics can be usefully projected onto that of a coarse-grained field, and (3) an atomistic demonstration that dewetting is key to the dynamics of hydrophobic collapse.

\section{Atomistic model, coarse-graining, and the string method}

\subsection{The system}
We consider the atomistic version of the hydrated hydrophobic chain
studied by ten Wolde and Chandler.\cite{ten02} The
chain is composed of twelve spherical hydrophobes, each of diameter
$7.2$ \AA ~and mass $70$ amu.  These are ideal hydrophobes, as they exert purely
repulsive interactions upon the oxygen atoms of
the water molecules.  Consecutive hydrophobes in the unbranched chain interact via harmonic
bonds, and the chain is made semi-rigid by a potential energy term
that penalizes its curvature. The chain is hydrated with
approximately $34000$ rigid water molecules in an orthorhombic
simulation box with periodic boundary conditions.  All simulations
were performed at $300$~K.  Full details of the system are provided in
the Supporting Information Sec. A.

In describing hydrophobic collapse, it is necessary to
choose an appropriate thermodynamic ensemble for the simulations.
Nanometer-scale fluctuations in solvent density are suspected to play
a key role in these dynamics.  Use of the NVT ensemble
(with a 1~g/cm$^3$ density of water) might suppress these density
fluctuations and bias the calculated mechanism.  
We avoid this problem with a simple technique that is based on the fact that
under ambient conditions, liquid water is very close to phase
coexistence.  Indeed, it is this proximity that leads to the
possibility of large lengthscale hydrophobicity.\cite{Cha05,Lum99}  By placing a fixed
number of water molecules at $300$ K in a volume that corresponds to
an average density of less than 1~g/cm$^3$, we obtain a fraction
of the system at the density of water vapor and the majority at a
density of bulk water. Since we are not concerned with solvent
fluctuations on macroscopic lengthscales, the difference between
simulating bulk water at its own vapor pressure compared to
atmospheric pressure is completely negligible.  This strategy has been previously
employed to study the dewetting transition between solvophobic surfaces.\cite{Bol00,Hua05}
To ensure that the liquid-vapor interface remains both flat and well-distanced
from the location of the chain, we repel particles from a thin layer at the top edge of the simulation box, as is discussed in Supporting Information Sec. A.

\subsection{Coarse-graining}

Throughout this study, we simulate the hydrated chain using
atomistic MD, but we employ the string method
using collective variables that describe the solvent density field.
A coarse-graining algorithm is developed to connect these atomistic and collective variable representations of the solvent.
Following ten Wolde and Chandler,\cite{ten02}
 the simulation box is partitioned into  a three-dimensional
lattice (48$\times$48$\times$56) of cubic cells with sidelength $l = 2.1$ \AA.
We label the cells with the vector ${\bf k}=(k_{\rm x},k_{\rm y},k_{\rm z})$, where each $k_{\zeta}$ takes on integer values bounded as follows: $1\le k_{\rm x} \le 48$, $1\le k_{\rm y} \le 48$, and $1\le k_{\rm z} \le 56$.
On this grid, the molecular density, $\rho({\bf r})$, is coarse grained into the field $P_{\bf k}$, where
\begin{equation}
P_{\bf k} = \int\textrm{d}{\bf r}\ \rho({\bf r})\sum_{\zeta=x,y,z}\Phi_{k_{\zeta}}({\bf r}\cdot {\bf 1}_{\zeta}).
\label{CGdensity}
\end{equation}
Here, the integral extends over the volume of the system, ${\bf 1}_{\zeta}$ denotes the unit vector in the ${\zeta}^{\rm th}$ Cartesian direction, and the coarse graining function, $\Phi_{k}(x)$, must be normalized, i.e., $1=\sum_k\Phi_{k}(x)$.
The particular function that we have chosen to use is
\begin{eqnarray}
\Phi_{k}(x)&=&\left.\int\textrm{d}y\ \phi(x-y)\ \right[h_k(x)h_k(y) \\
&& \left.+ h_{k+1}(x)\sum_{j\le k}h_j(y) + h_{k-1}(x)\sum_{j\ge k}h_j(y)\right]\nonumber
\end{eqnarray}
where 
\begin{equation}
\phi(x)=(2\pi\sigma^2)^{-1/2}\ \textrm{exp}(-x^2/2\sigma^2),
\end{equation}
$\sigma=1 $\AA, and $h_k(x)$ is unity when $x$ is in the $k^{\rm th}$ interval and zero otherwise.  Since $\Phi_{k}(x)$ is normalized, $\sum_{\bf k} P_{\bf k}=N$ is the total number of water molecules.
In effect, this choice spreads the atomistic density field $\rho({\bf r})$ over the lengthscale $\sigma$ and bins it into a grid of lenthscale $l$ in such a way as to preserve normalization.
While we have found this choice of coarse graining function to be convenient, others are possible.

\begin{figure}[!tbp] 
      \hspace*{-0.5cm}\includegraphics[angle=90,width=9.5cm,clip=t]{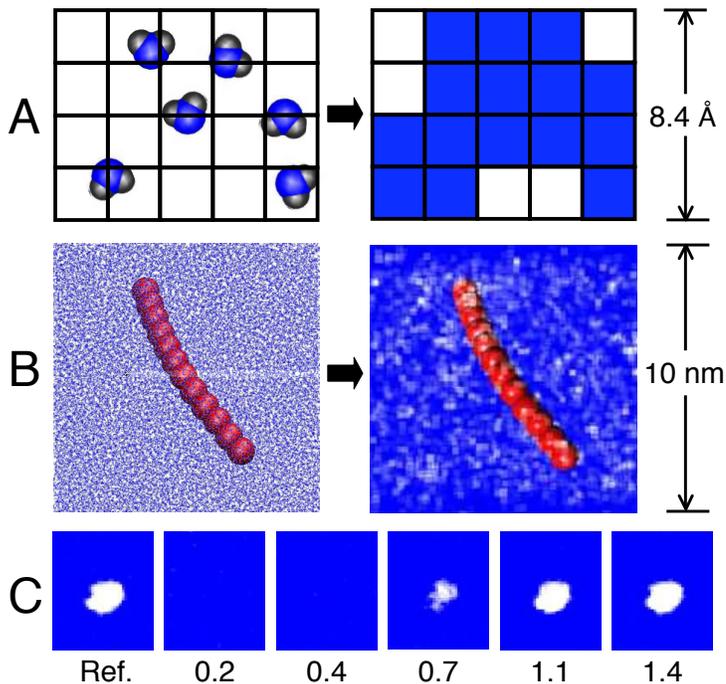}\\
   \vspace{-0.2cm}\mbox{}
 \caption[sqr]
 {\label{shmear} (A) The coarse-graining procedure is schematically shown to project the atomistic solvent density onto a discrete grid.
 (B) The same procedure is shown for an instantaneous solvent configuration of the actual system.  Grid cells containing less solvent density than $c/2$ are colored white; the remaining cells are left transparent against the blue background.  The hydrophobic chain is shown in red.
 (C) The potential in Eq. \ref{shmear4} is used to restrain atomistic solvent
   density to the reference distribution at the far left.  With
   larger $\kappa$, reported numerically in units
   of $2 k_BT/c^2$, the average atomic solvent density reproduces the reference distribution in
   detail (see text for notation).  }
\end{figure}

Figs. \ref{shmear}A and B illustrate the coarse-graining procedure.
In part A, the solvent is schematically shown before and after coarse-graining.  In part B, the same mapping is shown for the actual system considered herein.
Cells are shaded white when their solvent occupation, $P_{\bf k}$, is less
than half of its bulk average value of $\langle P\rangle_{\rm bulk}\equiv c=0.3$ molecules.  Small
local density fluctuations are seen throughout the simulation box, as is expected for an instantaneous solvent configuration.

In addition to visualizing solvent density, the coarse-graining
algorithm is useful for controlling the solvent density in MD simulations.  
For example, if it is desired that a particular cell ${\bf k}$ exhibit a solvent occupation $P^*_{\bf k}$, the following potential energy term can be used to derive the appropriate forces on the atoms in the simulation,
\begin{equation}
\frac12 \kappa(P_{\bf k}-P^*_{\bf k})^2,
\label{shmear4}
\end{equation}
where $\kappa$ is an appropriate force constant.
This technique is illustrated in Fig. \ref{shmear}C.
The left-most panel shows a density distribution that is exceedingly
unlikely for a simulation of ambient liquid water.  Panels to the
right show the average solvent density distribution from MD simulations that are restrained to this unlikely reference
distribution with increasing $\kappa$.
  Force constant values greater than $2 k_BT/c^2$,
in which the simulation incurs an energetic penalty of at least $k_BT$
for placing the average bulk density in a cell that is restrained to
be empty, ensure that the reference distribution is recovered in
detail.

\subsection{String method in collective variables}

We use the string method in collective variables to calculate the minimum free energy path (MFEP) for the collapse of the hydrated chain.\cite{E02,E05,Mar06}
To explain, we introduce some notation.  Let ${\bf x}=({\bf x}_{\rm c},{\bf w})$ be the position vector of length $n=3\times 12+3\times 3\times N$ for the atomistic representation of the entire system, where ${\bf x}_{\rm c}$ is the position vector for the atoms in the chain, and ${\bf w}$ is the position vector for the atoms in the water molecules.  Similarly, let ${\bf z}({\bf x})=({\bf x}_{\rm c},{\bf P})$ be the vector of length ${\cal N}=3\times 12+48\times 48\times 56$ for the collective variable representation of the system, where the elements of ${\bf P}$ are defined in Eq. \ref{CGdensity}.  

The MFEP, ${\bf z}^*(\alpha)$, is parameterized by the string coordinate $\alpha$, where $\alpha~=~0$ corresponds to the collapsed chain and $\alpha~=~1$ corresponds to the extended chain.  It obeys the condition
\begin{equation}
\frac{d z^*_i(\alpha)}{d\alpha} \quad\textrm{parallel to}\quad \sum_{j=1}^{\cal N} M_{ij}({\bf z}^*(\alpha))\frac{\partial F({\bf z}^*(\alpha))}{\partial z_j},
\label{MFEPcondition}
\end{equation}
where 
\begin{eqnarray}
F({\bf z})&=&-\beta^{-1}\textrm{ln} \left<\prod_{i=1}^{\cal N}\delta(z_i-z_i({\bf x}))\right>
\end{eqnarray}
 is the free energy surface defined in the collective variables, and
\begin{eqnarray}
M_{ij}({\bf z}) &=& \textrm{exp}[\beta F\left({\bf z}\right)]\\
& &\times\left<\sum_{k=1}^n m_k^{-1}
\frac{\partial z_i({\bf x})}{\partial x_k}
\frac{\partial z_j({\bf x})}{\partial x_k}
\prod_{i=1}^{\cal N}\delta(z_i-z_i({\bf x}))\right>.\nonumber
\end{eqnarray}
  Here, angle brackets indicate equilibrium expectation values, $\beta=(k_BT)^{-1}$ is the reciprocal temperature, and $m_k$ is the mass of the atom corresponding to coordinate $x_k$.

If the employed collective variables are adequate to describe the mechanism of the reaction (here, the hydrophobic collapse), then it can be shown that the MFEP is the path of maximum likelihood for reactive MD trajectories that are monitored in the collective variables.\cite{Mar06}
In the current application, we shall check the adequacy of the collective variables \textit{a posteriori} by running MD trajectories that are initiated from the presumed rate-limiting step along the MFEP (i.e. the configuration of maximum free energy) and verifying that these trajectories lead with approximately equal probability to either the collapsed or extended configurations of the chain (see Section \ref{qsection}).  

The string method yields the MFEP by evolving a parameterized curve (i.e. a string)
 according to the dynamics\cite{Mar06,E07}
 \begin{equation}
  \label{steepest1}
  \begin{aligned}
    \frac{\partial z^*_i(\alpha,t)}{\partial t} & =-\sum_{j=1}^N
    M_{ij}({\bf z}^*(\alpha,t))\frac{\partial
      F({\bf z}^*(\alpha,t))}{\partial z_j}\\
    & \quad + \lambda(\alpha,t) \frac{\partial z^*_i(\alpha,t)}{\partial
      \alpha}
  \end{aligned}
\end{equation}
where the term $ \lambda(\alpha,t) \partial z^*_i(\alpha,t)/\partial \alpha$ enforces the constraint that the string remain parameterized by normalized arc-length.
The endpoints of the string evolve by steepest decent on the
free energy surface,
\begin{equation}
\frac{\partial z^*_i(\alpha,t)}{\partial t}=-\frac{\partial F({\bf z}^*(\alpha,t))}{\partial z_i},
\label{steepest2}
\end{equation}
for $\alpha=0$ and $\alpha=1$.
These artificial dynamics of the string yield the MFEP, which satisfies Eq. \ref{MFEPcondition}.
  
  In practice, the string is discretized using $N_{\rm d}$ configurations of the system in the collective variable representation.  The dynamics in Eqs. \ref{steepest1} and \ref{steepest2} are then accomplished in a three-step cycle
  where (i) the endpoint configurations of the string are evolved according to Eq. \ref{steepest2}  and the rest of the configurations are evolved according to the first term in Eq. \ref{steepest1}, (ii) the string is (optionally) smoothed, and (iii) the string is reparameterized to maintain equidistance of the configurations in the discretization.  Step (i) requires evaluation of the  mean force elements
  $\partial F({\bf z})/\partial z_i$ and the tensor elements $M_{ij}({\bf z})$ at each configuration.
  These terms are obtained using restrained atomistic MD simulations of the sort illustrated for the solvent degrees of freedom in Fig. \ref{shmear}C.
The details of the string calculation are provided in Supporting Information Sec. B.
  
\section{Hydrophobic collapse of a hydrated chain}

\subsection{Minimum free energy path}

\begin{figure}[!tbp] 
\hspace*{-.5cm}\includegraphics[angle=0,width=9.5cm,clip=t]{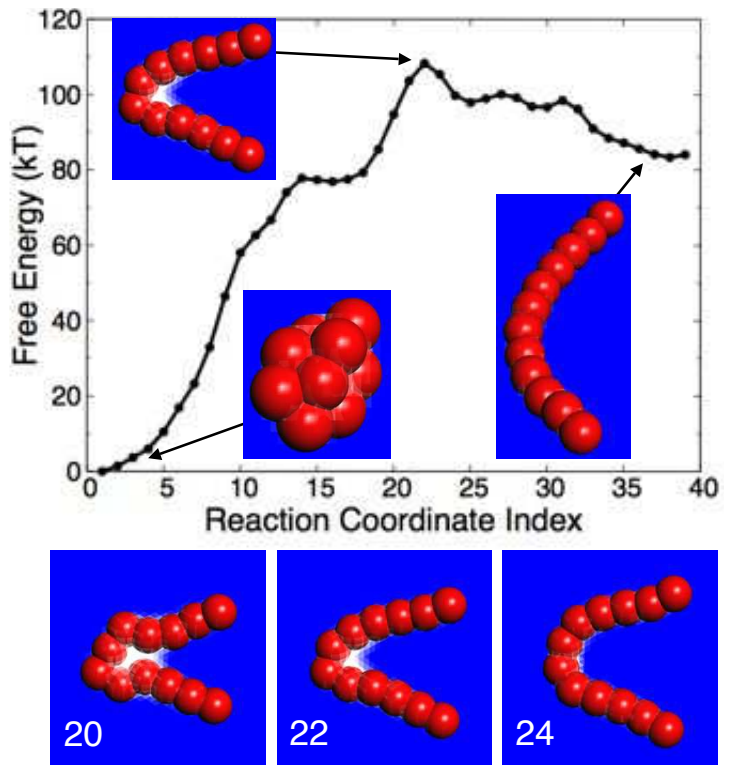}\\
   \vspace{-0.5cm}\mbox{}
 \caption[sqr]
 {\label{FEprofile} The minimum free energy path obtained using the
   string method.  Above, the free energy profile exhibits a single peak at configuration
   22.  Below, the configurations of the path in the vicinity of the
   free energy peak are shown with configuration numbers indicated in white text.
 }
\end{figure}

Fig. \ref{FEprofile} shows the MFEP for the hydrophobic collapse of the hydrated chain.  It is obtained using the string method in the collective variables for the chain atom positions and the grid-based solvent density field.  The string is discretized using $N_{\rm d}=40$
configurations of the system, and it is evolved using the steepest descent dynamics in Eqs. \ref{steepest1} and \ref{steepest2} to
yield the MFEP that satisfies Eq. \ref{MFEPcondition}.
The free energy profile is obtained by integrating the projection of the mean force along the MFEP, using
\begin{equation}
F^*(\alpha) = \int_0^\alpha \nabla F({\bf z}^*(\alpha'))\cdot d{\bf z}^*(\alpha').
\label{FEprofEQ1}
\end{equation}
Upon discretization, $s=N_d\times\alpha$ is the configuration number that indexes the MFEP.
The resolution of $F^*(\alpha)$ in Fig.  \ref{FEprofile} could be improved by employing a larger $N_{\rm d}$, but at larger computational cost.

To the extent that the collective variables employed in this study adequately describe the mechanism of hydrophobic collapse, the variable $s$ parameterizes the reaction coordinate for the collapse dynamics; this assumption is checked below with the use of straightforward MD simulations.
The statistical errors in the free
energy profile between consecutive configurations are approximately
the size of the plotted circles, and the small features in the profile
at configurations 27 and 31 are due to noise in the convergence of the
string calculation.
The lower panel of Fig. \ref{FEprofile} presents configurations
along the MFEP in the region of the free energy barrier.  As in Fig.
\ref{shmear}, lattice cells with less than half of the bulk solvent
occupation number fade to white.

The free energy profile in Fig. \ref{FEprofile} is dominated by a
single barrier at configuration 22, where a liquid-vapor interface is
formed at a bend in the hydrophobic chain.  The sharply curved chain
geometry presents an extended hydrophobic surface to molecules located
in the crook of the bend, an environment that is analogous to that
experienced by water trapped between hydrophobic plates and known to
stabilize large-lengthscale solvent density
fluctuations.\cite{Lum99,Luz00a,Luz00b} 
The barrier in the calculated free energy profile clearly coincides
with a collective motion in the solvent variables.

\subsection{The solvation free energy}

\begin{figure}[!tbp] 
\vspace{-.75cm} \hspace*{0cm} \includegraphics[angle=0,width=7.2cm,clip=t]{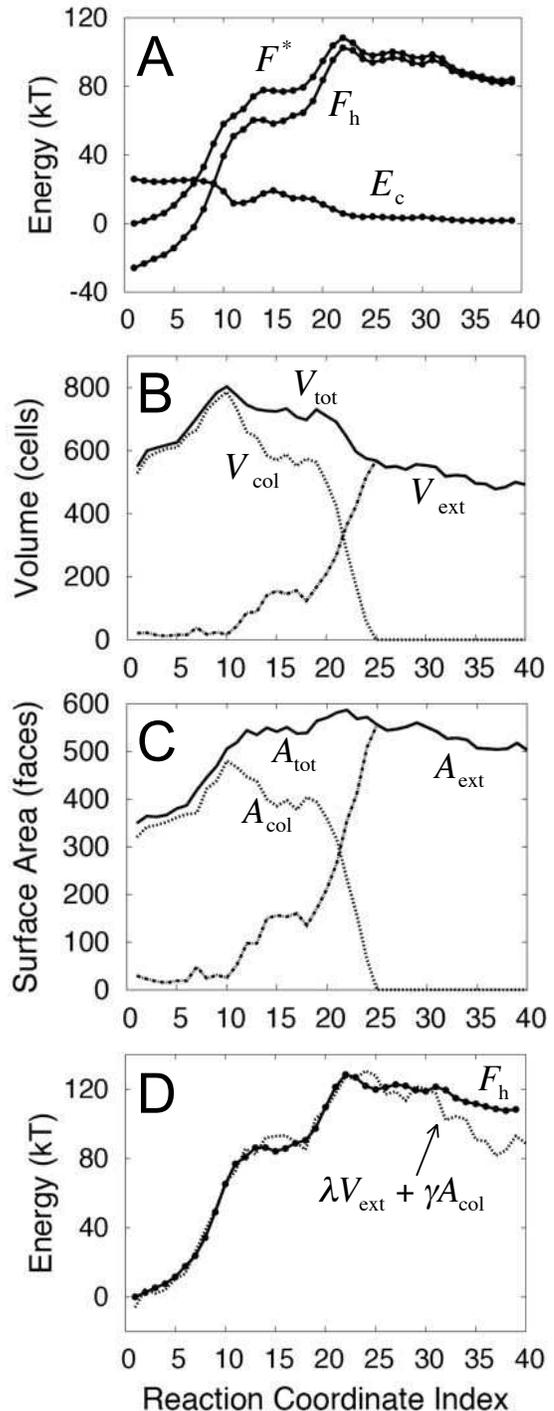}\\
   \vspace{0.0cm}\mbox{}
 \caption[sqr]
 {\label{theory} 
   A simple theory describes the solvation free energy
   of irregular hydrophobic solute geometries.  \textbf{(A)} The
   calculated free energy profile $F^*$ is decomposed into contributions
   from the chain configurational potential energy $E_{\rm c}$ and
   the free energy of solvation $F_{\rm h}$.  \textbf{(B)} The volume
   of the hydrated chain is parsed into large-lengthscale
   ($V_{\rm col}$) and small-lengthscale ($V_{\rm ext}$)
   components.  \textbf{(C)} The surface area of the chain is
   similarly parsed.  \textbf{(D)} The dotted line indicates the solvation
   free energy obtained using a relation (Eq. \ref{fsolvtheory}) that
   is based on the volume-scaling of the small-lengthscale solvation
   free energy and the surface area-scaling of the large-lengthscale
   solvation free energy.  The solvation free energy
   obtained from simulation (solid line) is included for comparison.}
\end{figure}

A simple theory can be constructed to understand the contributions to
the free energy profile in Fig. \ref{FEprofile} and to test the
validity of coarse-graining solvent interactions.  Noting that our choice of collective variables for
 the string calculation neglects the configurational
entropy of the chain, the free energy profile $F$ can be decomposed
into the configurational potential energy of the chain $E_{c}$
and the free energy of hydrating the chain $F_{h}$,
\begin{equation}
F^*(\alpha) = E_{\rm c}(\alpha)+F_{\rm h}(\alpha).
\label{decompose}
\end{equation}
The term $E_{\rm c}(\alpha)$ is easily evaluated from the chain potential
energy term, yielding the components of the free energy profile
shown in Fig. \ref{theory}A.

We focus our attention on $F_{\rm h}$.
The solvation free energy for hydrophobes of idealized
geometry is well understood.
For small-lengthscale hydrophobes ($<$ 1 nm), this energy  scales linearly with the solute
volume, whereas for large-lengthscale hydrophobes, it scales
linearly with the solute surface area.\cite{Lum99,Hua01} 
But theoretical prediction of the solvation free energy for more
complicated solute geometries is not necessarily trivial.  For
example, depending on its configuration, some parts of the chain
considered here might be solvated like a large-lengthscale hydrophobe,
whereas other parts might behave like a small-lengthscale hydrophobe.

We have developed a technique for
parsing a general hydrophobic solute into components that belong to
either the small-lengthscale or large-lengthscale regimes.
We define the
solvent-depleted volume as the continuous set of lattice
cells that are, on average, occupied by less than 50\% of the average bulk solvent value
 (this set includes both the
volume that is excluded by the hard-sphere-like interactions
between the solute and solvent interactions, as well as any additional
volume in the vicinity of the solute that is not substantially
occupied by the solvent).  We also introduce a probe volume that is a cubic $4\times 4\times 4$ set of
cells, which is approximately the size at which the water solvation
structure changes from the small-lengthscale to the large-lengthscale
regime.

The probe volume is used to determine whether a given section of the
solvent-depleted volume is in the small- or large-lengthscale regime.
This is done by moving the probe volume throughout the simulation box
and determining whether it can be fit into different portions of the
solvent-depleted volume.  If, at a given position, the vast majority of
this probe volume (specifically 59 out of 64 cells) fits within the
solvent-depleted volume, then the cells in that portion of the solvent-depleted
volume are included in the large-lengthscale component, $V_{\rm col}$.  Portions of the
solvent-depleted volume that do not meet this criterion for any
position of the probe volume 
are included in the small-lengthscale
component, $V_{\rm ext}$.  The volumes $V_{\rm col}$ and  $V_{\rm ext}$, which are plotted in Fig. \ref{theory}B, primarily include contributions from the collapsed portions and extended portions of the chain, respectively.  The total volume $V_{\rm tot}$ is obtained by counting the total number of cells in
the solvent-depleted volume, such that $V_{\rm tot}=V_{\rm ext}+V_{\rm col}$.

The total surface area, $A_{\rm tot}$, and its large-lengthscale component, $A_{\rm col}$, were obtained by counting
the number of external faces on the cells that comprise $V_{\rm tot}$ and $V_{\rm col}$, respectively.
  The surface area of the small-lengthscale component was
obtained using $A_{\rm ext}=A_{\rm tot}-A_{\rm col}.$ 
To account for the fact that we are calculating the area of a smooth
surface by projecting it onto a cubic lattice, each surface area term was also multiplied by a $2/3$, a correction factor that is exact for an infinitely large sphere.  The surface area components are plotted in Fig. \ref{theory}C.

We use the parsed components of the solute volume and surface area
to estimate the solvation free energy for
the chain along the minimum free energy path with the relationship
\begin{equation}
F_{\rm h}(\alpha) = \lambda V_{\rm ext}(\alpha) + \gamma A_{\rm col}(\alpha).
\label{fsolvtheory}
\end{equation}
The coefficients $\lambda$ and $\gamma$ are, respectively, the prefactors for the
linear scaling of the solvation free energy in the small-lengthscale
and large-lengthscale regimes, as obtained by calculations on a
spherical hydrophobic solute.\cite{Hua01}  Their values are taken to be 
$\lambda=8$ mJ/(m$^2$\AA) and $\gamma=45$ mJ/m$^2$.

The results of this
calculation are
presented in Fig. \ref{theory}D. 
The dotted line indicates our theoretical
estimate of the solvation free energy using Eq. \ref{fsolvtheory}, and the solid line indicates the
data from the atomistic simulations (from Fig.
\ref{theory}A).  
The agreement is very good, and as is shown in
Supporting Information Sec. C, it is better than can be obtained from
solvation free energy estimates that exclusively consider either the
solute volume or the solute surface area.  This result demonstrates that atomistic solvent energetics can be quantitatively modeled using a grid-based solvent density field.  
Furthermore, we note in
Fig. \ref{theory}B and C that the collapsing chain first develops
large-lengthscale components in the vicinity of the peak in the free
energy profile, which suggests that this peak coincides
with the crossover from small- to large-lengthscale solvation (i.e. dewetting).
This observation is consistent with the collective solvent density motion that is observed near the
free energy barrier in Fig. \ref{FEprofile}.

\subsection{The committor function and a proof of principle for coarse graining}
\label{qsection}

The various assumptions that are employed in our implementation of the
string method, including our choice of collective variables and the
corresponding neglect of the chain configurational entropy, raise
the possibility that the barrier in the
free energy profile in Fig.~\ref{FEprofile} does not
correspond to the dynamical bottleneck for hydrophobic collapse.  This is a general
concern in trying to relate free energy calculations to dynamical quantities, such as the reaction mechanism and the reaction rate. 
To confirm that the calculated free energy profile is dynamically relevant, we evaluate the committor function at several configurations along the MFEP.  The committor function reports the relative probability that MD trajectories that are initialized from a particular collective variable configuration first proceed to the extended configurations of
the chain, as opposed to the collapsed configurations.
For initial collective variable configurations that coincide with the true dynamical bottleneck, the committor function assumes a value of exactly $0.5$.

\begin{figure}[!tbp] 
  \hspace*{-.5cm}\includegraphics[angle=0,width=8.0cm,clip=t]{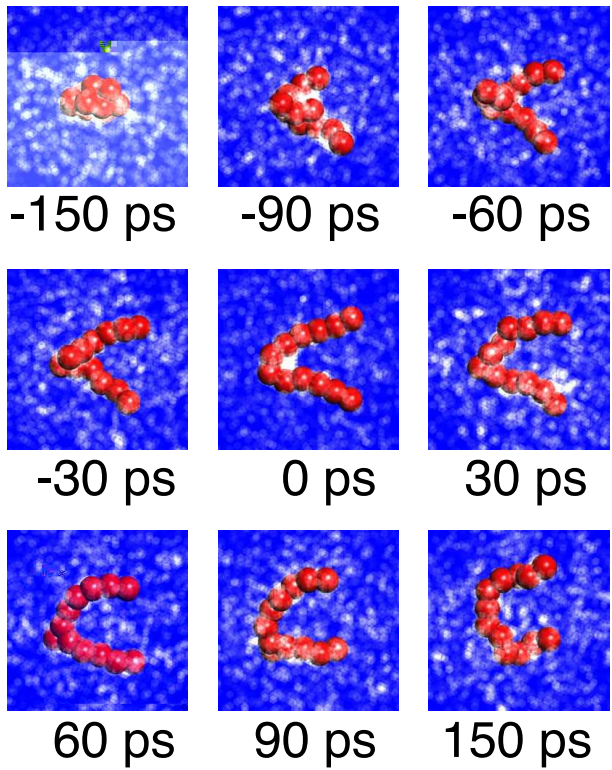}\\
   \vspace{-0.4cm}\mbox{}
 \caption[sqr]
 {\label{mdtraj} Snapshots from unrestrained MD trajectories that are initiated
 from the calculated free energy barrier at configuration 22.  The instantaneous atomistic configurations are visualized in the collective variable representation,
using the same technique that was introduced in Figs. \ref{shmear} A and B.}
\end{figure}

We first evaluate the committor function at configuration 22, which corresponds to the peak in the calculated free energy profile in Fig. \ref{FEprofile}.
The committor function is obtained by performing straightforward MD trajectories from initial conditions that are consistent with this collective variable configuration, and then tallying the fraction of those trajectories whose endpoints are ``extended,'' as opposed to ``collapsed''.
The initial conditions for these trajectories are obtained from a 20 ps MD trajectory that is restrained to the collective variables for configuration 22; the atomistic coordinates of the restrained trajectory are recorded every picosecond.
From each set of atomistic coordinates, an unrestrained trajectory was run for 150 ps forwards and backwards
in time with the initial atomistic velocity vector drawn from the
Maxwell-Boltzmann distribution at $300$ K.   To decide whether a given unrestrained trajectory
terminates in either an extended or collapsed configuration of the chain, we employ an order parameter
based on the number of chain atoms that are within $1.25$ \AA
~of another chain atom to which it is not directly bonded.  If the number
of non-bonded contacts exceeds three, 
the chain is considered to be in
a collapsed configuration; otherwise it is considered extended.
This order parameter need only distinguish between
collapsed and extended configurations of the chain; it need not be
(and in fact is not) a good reaction coordinate.
Fig. \ref{mdtraj} illustrates representative forward and backwards unrestrained MD trajectories.

The committor function at configuration 22 is
$0.3\pm0.1$, suggesting that the calculated free energy barrier is
slightly biased towards the basin of stability for the collapse chain. 
However, this deviation from the ideal value of $0.5$ is only
marginally statistically significant, and we emphasize that the
committor function is exponentially sensitive in the region of the
dynamical bottleneck.  To illustrate this point, we repeat the evaluation
of the committor function at configuration 20, which is seen in Fig.
\ref{FEprofile} to be near the barrier peak but slightly closer to
the collapsed configurations of the chain, as well as at configuration 24, which is slightly closer to the extended configurations.  We find the
committor function at configuration 20 to be $0.00\pm0.05$ and the value at configuration 24 to be $0.90\pm0.05$.  Only very minor shifts along the MFEP dramatically change the value of the committor function.

Given the
extreme sensitivity of the commitor function near the dynamical
bottleneck, the fact that configuration 22 gives rise to a significant
fraction of trajectories that proceed to both the collapsed and
extended basins is very significant.
Furthermore, the direction in which the committor function changes as a function of the configuration number is as is expected in the vicinity of the dynamical bottleneck.
We conclude that the barrier in the calculated
free energy profile correctly characterizes the true free energy barrier - thus identifying the rate-limiting step for the collapse dynamics. 

The committor function calculations, which are based on straightforward MD simulations, indicate that our choice of collective variables provides a reasonable description of the mechanism of hydrophobic collapse.
This result yields atomistic support for the strategy of coarse-graining solvent dynamics.
Using (i) that it is straightforward\cite{Mar06} to obtain the stochastic dynamics in the collective variables for which the most likely reaction pathway is the same as the mechanism obtained using the string method
and (ii) that the mechanism for hydrophobic collapse obtained in the employed collective variables agrees with the mechanism obtained using atomistic dynamics, we obtain a
proof of principle that atomistic solvent dynamics can be usefully projected onto a coarse-grained field.

The coarse-grained dynamics obtained via this argument are\cite{Mar06}
\begin{eqnarray}
\bar{\gamma}\frac{\partial z(t)}{\partial t}&=&-\sum_{j=1}^{\cal N}
\left(M_{ij}({\bf z}(t))\frac{\partial F(z(t))}{\partial z_j}-\beta^{-1}\frac{\partial M_{ij}(z(t))}{\partial z_j}\right)\nonumber\\
&& +\sqrt{2\beta^{-1}\bar{\gamma}}\sum_{j=1}^{\cal N}M^{1/2}_{ij}(z(t))\eta_j(t),
\end{eqnarray}
where $\eta_i(t)$ is a white noise satisfying $\langle\eta_i(t)\eta_j(t)\rangle=\delta_{ij}\delta(t-t')$ and $t$ is an artificial time that is scaled by the friction coefficient $\bar{\gamma}$.
The computational feasibility of directly integrating these coarse-grained dynamics, however, hinges on the cost of calculating the mean force elements
  $\partial F({\bf z})/\partial z_j$ and the tensor elements $M_{ij}({\bf z})$ and $\partial M_{ij}({\bf z})/\partial z_j$, 
 since these terms are required at every coarse-grained timestep.
It is thus encouraging that the free energy surface in the collective variables can be quantitatively modeled in lieu of atomistic simulations, by using Eqs. \ref{decompose} and \ref{fsolvtheory}.
Similar approximations for the tensor elements might also be possible.

\subsection{The rate-limiting step}

Using the calculated MFEP and committor function values, we have established that the rate-limiting step for the hydrophobic collapse of the hydrated chain coincides with a collective solvent motion.  Using a simple analysis of the solvation free energy, we find that this collective solvent motion is consistent with lengthscale-dependent dewetting.
However, it remains to be shown whether the rate-limiting step involves performing work in the solvent, or in the chain, degrees of freedom.  This is an important distinction.
If the latter case is true, then dewetting merely accompanies hydrophobic collapse as a spectator.  But if the former case is true, then dewetting \textit{is} the rate-limiting step to hydrophobic collapse.

To address this issue, we again decompose the free energy profile,
 this time into contributions from work performed along the solvent and the chain collective variables.
The definition of the free energy profile in Eq. \ref{FEprofEQ1} can be written more explicitly as
\begin{eqnarray}
F^*(\alpha) &=& \int_0^\alpha \nabla_{\rm c}F({\bf z}^*(\alpha'))\cdot d{\bf x}_{\rm c}(\alpha')\nonumber\\
&+& \sum_{{\bf k}\in V}\int_0^\alpha \nabla_{P_{\bf k}}F({\bf z}^*(\alpha'))dP_{\bf k}(\alpha')
\label{FEprofEQ}
\end{eqnarray}
where $\nabla_{\rm c}F({\bf z}^*(\alpha'))$ is the vector of mean forces acting on the chain atom positions at configuration $s=N_{\rm d}\times\alpha$ along the MFEP, and
$\nabla_{P_{\bf k}}F({\bf z}^*(\alpha'))$
 is the corresponding mean force on the solvent collective variable $P_{\bf k}$.
The full free energy
profile is obtained by setting the volume $V$ equal to the entire
simulation box.  But to understand the role of solvent in hydrophobic
collapse, it is informative to calculate the free energy profile using various smaller solvent volumes.

\begin{figure} [!tbp] 
\hspace*{-0.4cm} \includegraphics[angle=90,width=9.cm]{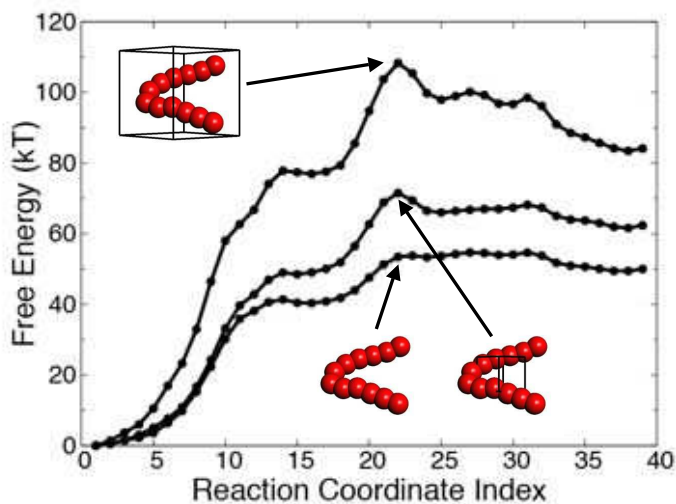}\\
  \vspace{-.0cm}\mbox{}
 \caption[sqr]
 {\label{FEbox} The free energy profile from Fig. \ref{FEprofile} is
   separated into contributions from the chain and solvent coordinates.
     In the bottom curve, only the first term in Eq. \ref{FEprofEQ}, which corresponds to work performed in the chain
     coordinates, is included in the free energy profile.  In the middle and top curves, solvent contributions are also included.
   The solvent volume included for each curve is indicated with a wire box.  Inclusion of larger solvent volumes does not
   substantially change the profile.}
\end{figure}

The bottom curve in Fig. \ref{FEbox} is obtained from Eq. \ref{FEprofEQ} by letting $V$ be the
empty set, thus eliminating the second term.
The middle curve is obtained by letting $V$ be the set of $8\times
8\times 8$ lattice cells at the middle of the simulation box, as is
indicated in the corresponding image, and the top curve is obtained by
letting $V$ be the middle set of $20\times 20\times 20$ lattice cells.
Consideration of larger sets of lattice cells does not further alter
the free energy profile, as is shown in Supporting Information Sec.
D.  The solvent variables for regions of space that are distant from
the collapsing chain remain constant along the path and do not
contribute to the free energy profile.

The bottom curve in Fig. \ref{FEbox} only shows the work performed on
the chain atom positions during hydrophobic collapse. Remarkably, it
lacks almost any free energy barrier, indicating that essentially no
work is performed in the chain degrees of freedom in crossing the
dynamical bottleneck.  Instead, we see that the barrier emerges only
upon inclusion of the work performed in the solvent collective
variables.  Performing the hydrophobic collapse involves traversing a
free energy barrier that exists only in the solvent coordinates.  This
figure shows that the dewetting transition not only accompanies
 the rate-limiting step for hydrophobic collapse, but that it
is the rate-limiting step.

\begin{acknowledgments} 
T.F.M. is supported by DOE grants CHE-0345280 and DE-FG03-87ER13793.  E.V.-E. acknowledges support from the Miller Institute at UC Berkeley, ONR grant N00014-04-1-0565, and NSF grants DMS02-09959 and DMS02-39625.  D.C acknowledges support from NSF grant CHE-0543158.  The authors also thank Giovanni Ciccotti, Mauro Ferrario, Ray Kapral, and Luca Maragliano for useful discussions and NERSC for computing resources.
\end{acknowledgments}

\newpage

\begin{center}
{\bf 
SUPPORTING INFORMATION
  }
  \end{center}
  \pagestyle{fancyplain}
\rhead[\fancyplain{try1}{try2}] {\fancyplain{try3}{\it Supporting Information}}
\lhead[\fancyplain{try1}{try2}] {\fancyplain{try3}{\it Miller, Vanden-Eijnden, and Chandler}}




\setcounter{subsection}{0}
\subsection{Details of the system}
We consider a hydrophobic chain in explicit solvent. The unbranched chain is composed of $N_{\rm s}=12$ spherical atoms of diameter $7.2$ \AA ~and mass $70.73$ amu.  The interactions between the chain atoms were treated with the purely repulsive WCA potential ($\sigma=6.4145 \AA, \epsilon = 15$ kJ/mol), which provides a slightly softened hard-sphere interaction that is convenient for molecular dynamics (MD) simulations.  Neighboring atoms are linked by harmonic bonds, using
\begin{equation}
V_{\rm bond}=\sum_{k=1}^{N_{\rm s}-1}\frac12 k_{\rm b} (l_{\rm b}-|{\bf r}_{k+1}-{\bf r}_k|)^2,
\end{equation}
where $|{\bf r}_{k+1}-{\bf r}_k|$ is the distance between chain atoms at positions ${\bf r}_k$ and ${\bf r}_{k+1}$, $l_{\rm b}=7.2$ \AA, and $k_{\rm b}=84.9102$ kJ/mol/\AA$^2$.  To add some rigidity to the chain, a harmonic potential is included to penalize its curvature,
\begin{equation}
V_{\rm angle}=\sum_{k=2}^{N_{\rm s}-1}\frac12 k_\phi \phi_k^2,
\end{equation}
where $\phi_k$ is the angle between neighboring bond vectors $({\bf r}_{k+1}-{\bf r}_k)$ and$({\bf r}_{k}-{\bf r}_{k-1})$, and $k_\phi=11.1407$ kJ/mol/radian$^2$.

The chain was solvated with $33,912$ water molecules in an orthorhombic simulation cell with periodic boundary conditions and dimensions of $99.5$ \AA $\times 99.5$ \AA $\times 116.1$ \AA.  Interactions among water molecules are described with the SPC/E rigid water potential,\cite{Ber87} and the chain atoms interact with the oxygen atoms in the water molecules via WCA repulsions\cite{WCA1,WCA2,WCA3,WCA4} ($\sigma=4.617$ \AA$, \epsilon = 10$ kJ/mol).  All MD simulations were performed at $300$ K using the Nos\'e-Hoover thermostat and a timestep of $2$ fs.\cite{Hoo85}
Constant pressure conditions were maintained in the simulations using the liquid-vapor coexistence technique described in the text.  To ensure that the liquid-vapor interface remains flat, as opposed to collapsing under the surface tension of the liquid, the solvent density restraint potential in Eq. \ref{shmear4} was applied to the top two layers of lattice cells in the simulation box with parameters $P^*_k=0$ and $\kappa = 100$ kJ/mol/molecule$^2$.
All MD calculations were performed using a modified version of the DL\_POLY\_3 molecular simulation package.\cite{DLPOLY3}

\subsection{String method in collective variables}
The string method~\cite{E02SA,E05SA,Mar06SA}is a technique for calculating the committor
function for a dynamical process.  The constant-value contours of the committor function are
 approximated by a collection of hyperplanes that are
 perpendicular to a given path (the string).
  This approximation can be justified within the framework of
  transition path theory (TPT),~\cite{Van06}
  which yields a variational criterion for the string such that the
  perpendicular hyperplanes optimally approximate the isocommittor surfaces.
  
   The string method in collective variables
   characterizes the committor function projected onto the space of collective variables.
   Provided that the collective variables adequately describe the reaction and that the reactive
  trajectories projected onto the space of collective variables remain
  confined to a relativelly narrow tube, the approximation is not only
  optimal but also accurate. In this case, it can be shown that the
  string coincides with the minimum free energy path (MFEP) in the space of
  collective variables, which is also the path of maximum likelihood
  for the recation monitored in these variables.
  
Unlike other free-energy
mapping techniques, the string method focuses only on a
linear subset of the space of collective variables.  It thus scales
independently of the dimensionality of the full free energy surface.  The
method does not require performance of long, reactive MD trajectories.
Instead, a double-ended approach is employed in which the path is
updated using short, restrained MD simulations.
  
A complete discussion of the implementation of the string method in collective variables can be
found in Ref. \onlinecite{Mar06SA}.  We represent the string in the
variables of the chain atom positions and the solvent cell
densities, as is explained in the primary text.  Data needed for the calculation of the minimum free
energy path were obtained from restrained MD
simulations, 
\begin{equation}
\frac{\partial F({\bf z})}{\partial z_i} = \lim_{\kappa\rightarrow\infty} \kappa\int_{\Re^n}
 (z_i-z_i({\bf x})) \rho_{\kappa,z}({\bf x}) d{\bf x}
 \label{data1}
 \end{equation}
and
\begin{equation}
M_{ij}({\bf z})\!=\!\lim_{\kappa\rightarrow\infty} \int_{\Re^n}\!\!
\sum_{k=1}^n
m_k^{-1}
\frac{\partial z_i({\bf x})}{\partial x_k}
\frac{\partial z_j({\bf x})}{\partial x_k}
\rho_{\kappa,z}({\bf x}) d{\bf x},
\label{data2}
\end{equation}
where 
\begin{eqnarray}
\rho_{\kappa,z}({\bf x})&=&\textrm{exp}(-\beta U_{\kappa,z}({\bf x}))/Z_{\kappa,z},\\
Z_{\kappa,z}&=&\int_{\Re^n}\textrm{exp}(-\beta U_{\kappa,z}({\bf x})) d{\bf x},\\
\textrm{and}\ \  U_{\kappa,z}({\bf x})&=&V({\bf x})+\frac{\kappa}{2}\sum_{i=1}^{\cal N}(z_i-z_i({\bf x}))^2.
\label{datalast}
\end{eqnarray}

The string is converged to the MFEP using Eqs. \ref{steepest1} and \ref{steepest2}.
Between timesteps of these dynamics, the string is smoothed and reparameterized.
The smoothing procedure is performed using
\begin{equation}
{\bf z}^{m,*}=(1-\eta){\bf z}^{m}+\frac{\eta}{2}({\bf z}^{m-1}+{\bf z}^{m+1}),
\end{equation}
where ${\bf z}^{m,*}$ and ${\bf z}^{m}$ indicate the $m^{\rm th}$ configuration of the smoothed and unsmoothed string, respectively.  The parameter $\eta=0.05$ was chosen to be sufficiently small ($O(N_{\rm d}^{-1})$, as is discussed in Ref. \onlinecite{Mar06SA}) that the smoothing procedure does not effect the accuracy of the calculated MFEP.
Reparameterization of the string was performed using the linear interpolation scheme described in 
Ref. \onlinecite{Mar06SA}.

The string was initialized from configurations of an MD trajectory in which
the hydrated chain was artificially extended from a collapsed configuration
with the aid of a greatly magnified force constant in the chain potential
energy term $V_{\rm angle}$.  In the first stage of the string calculation, the
string was discretized using $20$ configurations and evolved using large timesteps
and weak collective variable restraints. Specifically, we
employed chain atom restraint force constants of $2$ kJ/mol/\AA$^2$,
solvent restraints of $10$ kJ/mol/molecule$^2$, and a steepest descent
timestep of $300$ fs. Restrained MD trajectories of $10$ ps were
performed during this stage.  This first stage of the string calculation included
nine steepest descent timesteps.  

In a second stage of the string calculation, the number of images used to
represent the path was increased to $40$, the chain atom restraints were
increased to $5$ kJ/mol/\AA$^2$, the solvent restraints were increased
to $40$ kJ/mol/molecule$^2$, the steepest descent timestep was
decreased to $40$ fs, and the restrained MD trajectory time was
increased to $20$ ps.  
The second
stage of the string calculation was terminated after six optimization
steps, at which point reasonable convergence, as determined by
monitoring changes in the path and its corresponding free energy
profile, was obtained.  Throughout both stages of the string calculation,
the path endpoint image corresponding to the extended configuration
was (after local relaxation) held fixed, and the other endpoint
corresponding to the collapsed chain was allowed to relax on the
free energy surface according to Eq. \ref{steepest2}.
The final, converged string bears little resemblance to the initial guess.  

The string method is a
local, rather than global, optimization scheme.  Different minimized free energy 
paths might have been obtained from string calculations started with
different initial paths.  However, the simple topology of the chain,
as well as the unrestrained MD simulations presented in the paper,
suggest that the calculated MFEP is a reasonable characterization of
the reaction mechanism for the hydrophobic collapse of the hydrated chain.

All computations were performed in parallel using 2.2 GHz AMD Operton processors.  Each step in the first of the stage of the string calculation required $600$ CPU hours.  Each step in the second stage required $2400$ CPU hours.  Each evaluation of the committor function described in Sec. \ref{qsection} required $15,000$ CPU hours.

\subsection{Alternative models for the solvation free energy}

\begin{figure}[!tbp] 
  \hspace*{-.7cm}\includegraphics[angle=0,width=8.5cm,clip=t]{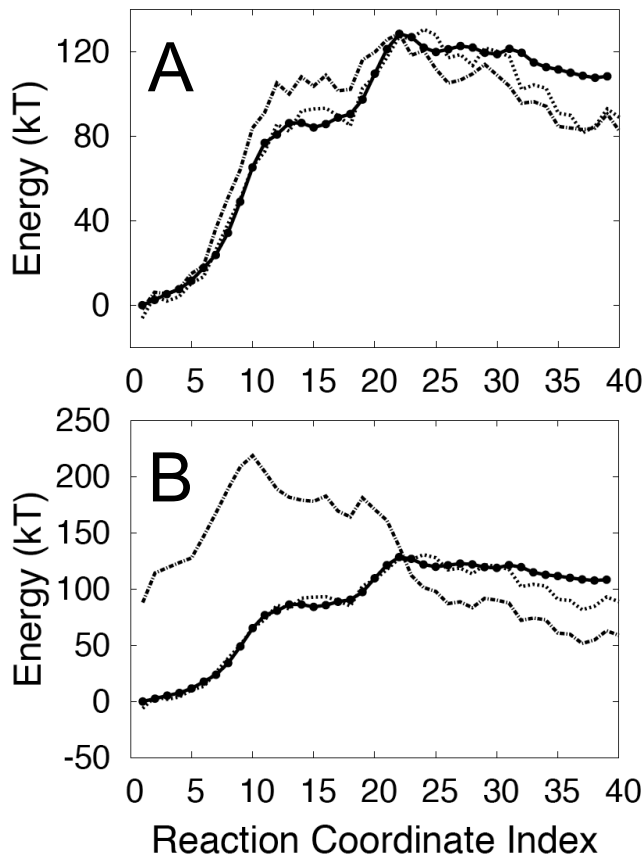}\\
   \vspace{-.2cm}\mbox{}
 \caption[sqr]
 {\label{otherfits} (Supporting Information) Alternative models for the  solvation free
   energy.  The profile obtained from simulation is
   shown as a solid line, and that obtained using Eq. \ref{fsolvtheory} is shown
   as a dotted line.  In Panel A, the dot-dashed line indicates the solvation free
   energy profile obtained from consideration of only the chain
   surface area, using Eq. \ref{SAonly}.  In Panel B, the dot-dashed line
   indicates the solvation free energy profile obtained from
   consideration of only the chain volume, using Eq. \ref{Vonly}.  }
\end{figure}

Even with the aide of a one-parameter fit, the solvation free energy
profiles estimated on the basis of only solvent-depleted surface area or only
solvent-depleted volume do not match the accuracy obtained using Eq. \ref{fsolvtheory} in the text.  In Fig.
\ref{otherfits}A, we see the profile (dot-dashed line) obtained by fitting
$\tilde{\gamma}$ in the expression
\begin{equation}
F_{\rm h}(\alpha) = \tilde{\gamma} A_{\rm tot}(\alpha).
\label{SAonly}
\end{equation}
The height of the shoulder at configurations 14-18 and the barrier
peak region do not reproduce the simulated results (solid line) as well as
Eq. \ref{fsolvtheory} (dotted line).  As is seen
Fig. \ref{otherfits}B, the solvation free energy profile
(dot-dashed) that is obtained using the expression
\begin{equation}
F_{\rm h}(\alpha) = \tilde{\lambda} V_{\rm tot}(\alpha)
\label{Vonly}
\end{equation}
is less accurate than that obtained using either Eqs. \ref{fsolvtheory}  or \ref{SAonly}.

\subsection{Convergence of the free energy profile with increasing solvent box sizes}

\begin{figure}[!tbp] 
 \hspace*{-1cm} \includegraphics[angle=90,width=8.5cm,clip=t]{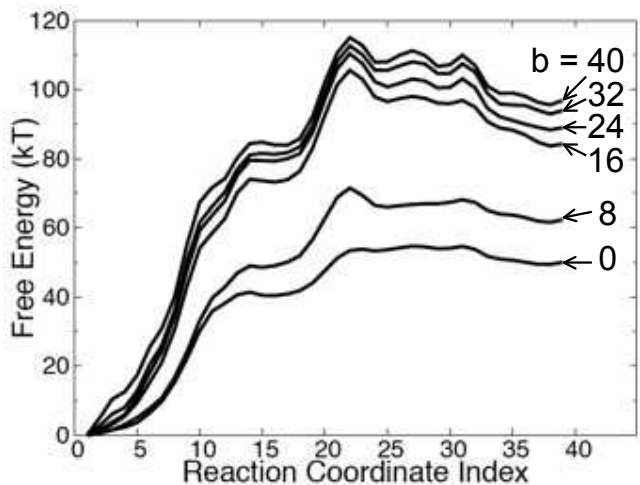}\\
   \vspace{-0cm}\mbox{}
 \caption[sqr]
 {\label{multi} (Supporting Information) Free energy profiles for the MFEP, calculated using Eq. \ref{FEprofEQ} with $V$ equal to the middle $b\times b\times b$ lattice cells of the simulation box.}
\end{figure}

We show that the free energy profile calculated using Eq. \ref{FEprofEQ} converges with respect to the solvent contribution.  The curves in Fig. \ref{multi} correspond to free energy profiles obtained with $V$ equal to the middle $b\times b\times b$ lattice cells of the simulation box.  The profile changes dramatically with $b$ until $V$ fully encompasses the region of the bending chain, where the dewetting transition occurs.   For $b>16$, no major changes are seen in the profile, since the added contributions are from solvent cells that are distant from the collapsing chain.  The small changes that are found with large $b$ are due to statistical noise.  This noise increases with larger $b$ since the number of included solvent cells increases as $b^3$.

\begin{thebibliography}{99}
\bibitem{E02} E W, Ren W, Vanden-Eijnden E (2002) \emph{Phys Rev B} 66:52301.
\bibitem{E05} E W, Ren W, Vanden-Eijnden E (2005) \emph{J Phys Chem B} 109:6688-6693.
\bibitem{Mar06} Maragliano L, Fischer A, Vanden-Eijnden E, Ciccotti G (2006) \emph{J Chem Phys} 125:024106.
\bibitem{ten02} ten Wolde PR, Chandler D (2002) \emph{Proc Natl Acad Sci USA} 99:6539-6543.
\bibitem{Kau59} Kauzmann W (1959) \emph{Adv Prot Chem} 14:1-63.
\bibitem{Saf94} Safron SA (1994) \textit{Statistical Thermodynamics of Surfaces, Interfaces and Membranes}, (Addison-Wesley, Reading), pp 1-95. 
\bibitem{Tan78} Tanford C (1978) \emph{Science} 200:1012-1018.
\bibitem{Cha05} Chandler D (2005) \emph{Nature} 437:640-647.
\bibitem{Hum00} Hummer G, Garde S, Garcia AE, Pratt LR (2000) \emph{Chem Phys} 258:349-370.
\bibitem{Hua03} Huang X, Margulis CJ, Berne BJ (2003) \emph{Proc Natl Acad Sci USA} 100:11953-11958.
\bibitem{Hua04} Huang Q, Ding S, Hua C-Y, Yang H-C, Chen C-L (2004) \emph{J Chem Phys} 121:1969-1977.
\bibitem{Liu05} Liu P, Huang X, Zhou R, Berne BJ (2005) \emph{Nature} 437:159-162.
\bibitem{Ath07} Athawale MV, Goel G, Ghosh T, Truskett TM, Garde S (2007) \emph{Proc Natl Acad Sci USA} 104:733-738.
\bibitem{Bol00} Bolhuis PG, Chandler D (2000) \emph{J Chem Phys} 113:8154-8160.
\bibitem{Hua05} Huang X, Zhou R, Berne BJ (2005) \emph{J Phys Chem B} 109:3546-3552.
\bibitem{E07} E W, Ren W, Vanden-Eijnden E, submitted.
\bibitem{Lum99} Lum K, Chandler D, Weeks JD (1999) \emph{J Phys Chem B} 103:4570-4577.
\bibitem{Luz00a} Luzar A, Leung K (2000) \emph{J Chem Phys} 113:5836-5844.
\bibitem{Luz00b} Leung K, Luzar A (2000) \emph{J Chem Phys} 113:5845-5852.
\bibitem{Hua01} Huang DM, Geissler PL, Chandler D (2001) \emph{J Phys Chem B} 105:6704-6709.
\end{thebibliography}

\begin{thebibliography}{99}
\bibitem{Ber87} Berendsen HJC, Grigera JR, Straatsma TP (1987) \emph{J Chem Phys} 91:6269-6271.
\bibitem{WCA1} Chandler D, Weeks JD (1970) \emph{Phys Rev Lett} 25:149-152.
\bibitem{WCA2} Weeks JD, Chandler D, Andersen HC (1971) \emph{J Chem Phys} 54:5237-5247.
\bibitem{WCA3} Weeks JD, Chandler D, Andersen HC (1971) \emph{J Chem Phys} 55:5422-5423.
\bibitem{WCA4} Andersen HC, Weeks JD, Chandler D (1971) \emph{Phys Rev A} 4:1597-1607.
\bibitem{Hoo85} Hoover WG (1985) \emph{Phys Rev A} 31:1695-1697.
\bibitem{DLPOLY3} Todorov I, Smith W (2004) \emph{Phil Trans R Soc Lond A} 362:1835-1852.
\bibitem{E02SA} E W, Ren W, Vanden-Eijnden E (2002) \emph{Phys Rev B} 66:52301.
\bibitem{E05SA} E W, Ren W, Vanden-Eijnden E (2005) \emph{J Phys Chem B} 109:6688-6693.
\bibitem{Mar06SA} Maragliano L, Fischer A, Vanden-Eijnden E, Ciccotti G (2006) \emph{J Chem Phys} 125:024106.
\bibitem{Van06} Vanden-Eijnden E (2006) in \emph{Computer Simulations in Condensed Matter: From Materials to Chemical Biology - Vol. 2}, eds Ferrario M, Ciccoti G, Binder K (Springer, LNP), 703:439-478.
\end{thebibliography}
\end{document}